# Attitudes of Children with Autism towards Robots: An Exploratory Study

**Andreia Costa**
University of Luxembourg
Esch-sur-Alzette, Luxembourg
andreia.pintocosta@uni.lu

**Tonie Schweich**
University of Luxembourg
Esch-sur-Alzette, Luxembourg
tonie.schweich.001@student.uni.lu

**Louise Charpiot**
University of Luxembourg
Esch-sur-Alzette, Luxembourg
louise.charpiot@uni.lu

**Georges Steffgen**
University of Luxembourg
Esch-sur-Alzette, Luxembourg
georges.steffgen@uni.lu

## Abstract
In this exploratory study we assessed how attitudes of children with autism spectrum disorder (ASD) towards robots together with children's autism-related social impairments are linked to indicators of children's preference of an interaction with a robot over an interaction with a person. We found that children with ASD have overall positive attitudes towards robots and that they often prefer interacting with a robot than with a person. Several of children's attitudes were linked to children's longer gazes towards a robot compared to a person. Autism-related social impairments were linked to more repetitive and stereotyped behaviors and to a shorter gaze duration in the interaction with the robot compared to the person. These preliminary results contribute to better understand factors that might help determine sub-groups of children with ASD for whom robots could be particularly useful.

## Introduction
Robots are rule-based and predictable systems, which can repeat patterns and can be organized and understood in a systematic way [1]. This corresponds to the characteristics of children with autism spectrum disorder (ASD), who have a desire for sameness, repetition, and an interest in inanimate objects [2].

Based on this principle, several robots have been developed to be used in interventions with children with ASD and robots have proved to be useful for these children [3]. However, not all children might equally benefit from interventions with robots and the benefits might be determined by children's attitudes towards robots and their social impairments. Therefore, it is important to study children's attitudes towards robots and children's autism-related social impairments that might determine their interaction with a robot.

To address this, in this exploratory study we assess the attitudes of children with ASD towards robots and then the link between attitudes and social impairments to indicators of preference of an interaction with a robot over an interaction with a person.

| #  | Age   | ASD Severity[a] | IQ[b] |
|----|-------|-----------------|-------|
| 1  | 9.22  | Moderate        | 97    |
| 2  | 8.21  | Severe          | 70    |
| 3  | 14.46 | Severe          | 107   |
| 4  | 8.22  | Severe          | 108   |
| 5  | 9.58  | Severe          | >85   |
| 6  | 6.04  | Mild            | 75    |
| 7  | 11.38 | Severe          | 60    |
| 8  | 13.42 | Severe          | 88    |
| 9  | 12.41 | Severe          | 70    |
| 10 | 9.04  | Severe          | 74    |

Table 1: Children's characteristics
[a]SRS-2 & DSM-5 (Clinical range compatible scales) [4]
[b]Wechsler Nonverbal Scale of Ability-WNV[5]

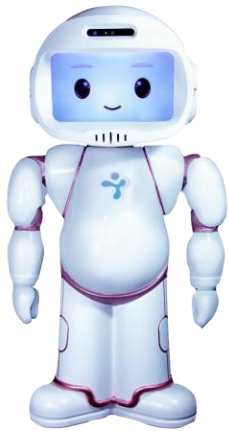

Figure 1: QTrobot

## Methods

*Participants*

A convenience sample of $N$=10 children (all boys) diagnosed with ASD, aged 6 to 14 years old ($M$=10.20; $SD$=2.64) participated in the study. Children's characteristics are described in Table 1. Due to Luxembourg's multilingualism, children's IQ was assessed with the Wechsler Nonverbal Scale of Ability (WNV [5]). Participants were part of a larger study on the validation of a robot-mediated emotional ability training for children with ASD.

*Instruments and Material*

CHILDREN'S ATTITUDES TOWARDS ROBOTS
Children's attitudes towards robots were assessed through 12 dual attitudes cards towards robots including attitudes such as robots' nature (evil or nice) distance from a robot, and preference of interaction. Cards were presented with a short explanation of each item insuring all children understood it. Higher scores indicate more positive attitudes towards robots.

CHILDREN'S AUTISM-RELATED SOCIAL IMPAIRMENTS
Children's autism-related social impairments were assessed through the Social Responsiveness Scale 2 (SRS-2; [4]). The SRS-2 has a composite scale score indicating severity and five subscales: social awareness, social cognition, social communication, social motivation, and repetitive and stereotyped behaviors. Higher scores indicate more difficulties.

ROBOT
The robot used to assess children's interactions, QTrobot (luxai.com; see Fig. 1), is a child-sized plastic bodied humanoid robot used in other recent applications for children with ASD [6, 7]. QTrobot has an expressive social appearance and its screen allows the presentation of animated faces. It has 12 degrees of freedom to present upper-body gestures.

CHILDREN'S INTERACTION: ROBOT VS PERSON
To compare children's interaction with a robot and with a human, we assessed children's gazes, imitation, and repetitive and stereotyped behaviors during an interaction with QTrobot and an equivalent interaction with a person. Details of this procedure and comparisons have been reported elsewhere [7]. Interactions lasted 1.8 to 4.2min. (QTrobot: $M$=2.98; $SD$=0.43; Person: $M$=2.46, $SD$=0.71). To measure the difference of children's interactions we subtracted the interaction with the person to the interaction with QTrobot in the different variables (Δ interaction).

*Procedure*

Parents read and signed informed consent forms for participation and data collection and the study was reviewed and approved by the University of Luxembourg's ethics review panel (approval number: ERP17-017-SAR-A). The study took place in one 2-hour long visit. During the visit, parents were requested to fill out questionnaires concerning their children. During that time, children were invited to a room where a researcher interacted with the child. After that, children's IQ, attitudes towards robots, and emotional abilities were assessed. At the end, children were invited to another room where they interacted with QTrobot. All children followed the same procedure and no counterbalancing of presentation order was done.

## Results

*Children's attitudes towards robots*
The frequencies of children's attitudes towards robots are displayed in Table 2. For each item, children indicated their preference or attitude. Frequencies on the left-hand side of the table are indicative of a preference or attitude towards robots; frequencies on the right-hand side of the table are indicative of a preference or attitude against robots. Missing frequencies (not adding up to 10) indicate the child had no preference or that both were true. Overall, 59% of the choices were towards robots, 32% against, and 9% indicated no preference.

| | |
|---|---|
| Robots are intelligent 8 | Robots are not intelligent 2 |
| Robot-like appearance 6 | Human-like appearance 4 |
| Robots are nice 8 | Robots are evil 0 |
| Prefer being close to a robot 5 | Prefer being far from a robot 4 |
| Robots do good things 8 | Robots do bad things 0 |
| Robots make me happy 7 | Robots make me scared 2 |
| Prefer robot as teacher 7 | Prefer person as teacher 3 |
| Prefer to work with a robot 4 | Prefer to work with a child 5 |
| Prefer to play with a robot 3 | Prefer to play with a child 5 |
| Prefer being friends with a robot 5 | Prefer being friends with a child 4 |
| Prefer to talk to a robot 4 | Prefer to talk to a child 5 |
| Prefer to tell a secret to a robot 6 | Prefer to tell a secret to a child 4 |

Table 2: Frequency of children's attitudes towards robots.

*Children's autism-related social impairments*
Children's social impairment scores measured by the SRS-2 [4] are displayed in Table 3. In the composite scale, most children scored on the severe range of the scale, indicating deficiencies in reciprocal behavior that lead to severe interferences with everyday social interactions [4]. Difficulties on the different subscales ranged from moderate to severe.

*Children's interaction: Robot vs Person*
The differences on gazes, imitation, and repetitive and stereotyped behaviors between children's interaction with QTrobot and with a person are displayed in Table 4. On average, children had more gazes per minute towards the person than towards the robot. However, the gaze average duration (in seconds) was longer for the robot than for the person. These two results indicate that children diverted their gaze from the person more often and looked longer at QTrobot. Additionally, the percentage of time looking at the interaction partner (% gaze duration) was longer in the robot condition than in the person condition. In terms of imitations, children imitated slightly more the robot than the person. A behavior was considered repetitive and stereotyped if the same behavior occurred at least 3 consecutive times. Children had more repetitive and stereotyped behaviors in the presence of the person than in the presence of the robot (#RSB per min.). Furthermore, when children engaged in these behaviors, the rhythm was faster in the person condition than in the robot condition (RSB rhythm).

*Relation between attitudes and social impairments to children's interaction with a robot over a person*
Bivariate correlations between children's attitudes towards robots and social impairments (SRS-2) and their interaction with QTrobot over a person are displayed in Table 5. Regarding children's attitudes towards robots, a preference to work and play with a robot were correlated to more gazes per minute to the robot than to the person and a positive attitude towards robots' nature and aims were correlated to longer gazes towards QTrobot and to a greater percentage of time looking at QTrobot compared to a person. Children's total SRS-2 score and increased difficulties in most subscales (all except social cognition) were linked to increased repetitive and stereotyped behaviors with QTrobot than with a person. Social communication difficulties were linked to more time spent looking at the person and a higher score in repetitive and stereotyped behaviors were linked to an increased amount of gazes towards the person.

## Discussion
The first aim of this study was to assess attitudes of children with ASD towards robots. The second aim was to explore how these attitudes as well as children's autism-related social impairments are related to children's interaction with a robot. We found that overall, children with ASD have more positive attitudes towards robots than against and that in several situations they prefer to interact with a robot than with a human. Furthermore, we found that children's preference to work and play with a robot over a human

|  | M (SD) |
|---|---|
| SRS - Total | 102 (20.69) |
| Social awareness | 11.90 (3.00) |
| Social cognition | 17.90 (3.73) |
| Social communication | 33.20 (7.51) |
| Social motivation | 16.60 (4.67) |
| Repetitive and stereotyped behaviors | 22.40 (5.91) |

Table 3: Children's scores in the SRS-2

|  | Δ M (SD) |
|---|---|
| # Gazes per min. | -1.48 (4.18) |
| Gaze avg. duration (sec) | 2.50 (2.80) |
| % Gaze duration | 31.21 (22.59) |
| # Imitations | 0.1 (0.32) |
| # RSB per min. | -2.83 (4.66) |
| # RSB rhythm | -9.92 (16.20) |

Table 4: Difference between children's interaction with QTrobot minus children's interaction with a person

|  | Δ # Gazes per min. | Δ Gaze. duration (sec.) | Δ % Gaze duration | Δ # Imitations | Δ # RSB per min. | Δ RSB rhythm |
|---|---|---|---|---|---|---|
| **Attitudes** | | | | | | |
| Intelligence | .298 | -.549 | -.288 | .167 | -.216 | -.223 |
| Appearance | -.131 | .447 | .383 | .272 | .149 | -.117 |
| Nature | -.045 | **.635*** | **.754*** | .167 | -.325 | -.335 |
| Proximity | -.042 | -.094 | -.047 | .318 | .407 | .017 |
| Aim | -.045 | **.635*** | **.754*** | .167 | -.325 | -.335 |
| Emotions | -.277 | .155 | -.266 | -.620 | .603 | .571 |
| School | .073 | .062 | .323 | .545 | .332 | .186 |
| Work | **.663*** | .026 | .512 | -.318 | -.435 | -.283 |
| Play | **.655*** | -.153 | .115 | -.306 | -.542 | -.375 |
| Friendship | .328 | -.104 | .089 | .318 | -.434 | -.413 |
| Conversation | .446 | -.031 | .354 | .389 | -.574 | -.561 |
| Secret | .116 | -.268 | -.608 | -.408 | .287 | .160 |
| **SRS - Total** | -.469 | -.208 | -.527 | .102 | **.884**** | **.666*** |
| Social awareness | .083 | .212 | -.152 | .129 | .589 | .389 |
| Social cognition | -.479 | -.330 | -.560 | .198 | **.779**** | **.739*** |
| Social communication | -.141 | -.538 | **-.672*** | -.009 | **.741*** | .574 |
| Social motivation | -.230 | -.313 | -.397 | .030 | **.770**** | .510 |
| Repetitive and stereotyped beh. | **-.689*** | .302 | -.248 | .155 | **.756*** | .537 |

Table 5. Bivariate correlations between children's attitudes, social impairments and difference of interaction with a robot and with a person (robot-person); *p<.05; **p<.01

was linked to longer gazes towards the robot than towards the person. Finally, social impairments were linked to more repetitive and stereotyped behaviors during the interaction with the robot and social communication difficulties and repetitive and stereotyped behaviors to less gazes towards the robot compared to the person.

These preliminary results contribute to understand the attitudes of children with ASD towards robots and factors that may be linked to the interaction of children with ASD with robots. The results based on our small sample indicate that children who already have positive attitudes towards robots look longer to a robot than to a person. However, those children who have more autism-related social impairments, particularly in terms of social communication and repetitive and stereotyped behaviors, might benefit less from an interaction with a robot compared to one with a person. Even though these results are preliminary and generalization cannot be guaranteed, the present results contribute to identify groups of children with ASD for whom interventions with robots might be more beneficial.